# Numerical Study on Mechanism of Multiple Rings Formation

Ping Lu, Yonghua Yan, Chaoqun Liu
UNIVERSITY OF TEXAS AT ARLINGTON, ARLINGTON, TX 76019, USA
CLIU@UTA.EDU

## ABSTRACT

In this paper, the flow around each ring-like vortex is investigated by high order DNS including first sweep, first ejection, second sweep, second ejection, positive spike, momentum deficit, vortex shape, vortex location, strength of sweeps, etc. Meanwhile, the mechanism about formation of momentum deficit is deeply studied. A new mechanism on how the multiple rings are formed one by one found both by experiment and by DNS in late boundary layer transition is presented. It also reveals that the relation between streamwise vortex and spanvise vertex rings, and how the vorticity is transferred between them.

## Nomenclature

$M_\infty$ = Mach number  
$Re$ = Reynolds number  
$\delta_{in}$ = inflow displacement thickness  
$T_w$ = wall temperature  
$T_\infty$ = free stream temperature  
$Lz_{in}$ = height at inflow boundary  
$Lz_{out}$ = height at outflow boundary  
$Lx$ = length of computational domain along x direction  
$Ly$ = length of computational domain along y direction  
$x_{in}$ = distance between leading edge of flat plate and upstream boundary of computational domain  
$A_{2d}$ = amplitude of 2D inlet disturbance  
$A_{3d}$ = amplitude of 3D inlet disturbance  
$\omega$ = frequency of inlet disturbance  
$\alpha_{2d}, \alpha_{3d}$ = two and three dimensional streamwise wave number of inlet disturbance  
$\beta$ = spanwise wave number of inlet disturbance  
$R$ = ideal gas constant  
$\gamma$ = ratio of specific heats  
$\mu_\infty$ = viscosity  

## I. Introduction

The formation of ring-like vortices chains is a physical phenomenon which has been observed by experimental work (Figure1) by Lee C B & Li R Q, 2007 and by Guo et al (2010). Kachanov, Rist and their group have done a great pioneering work (rist et al, 1995). Boroduln et al (2002) believed that the mechanism of formation multiple rings is controlled by Crow theory (Crow, 1970). Later, this hypothesis is supported by the computation of Bake and his co-authors (Bake et al 2002). Bake et al (2002) believe the multiple ring structure is formed by leading vortex breakdown and reconnection and so on. According to Kloker et al (2011), their DNS results "contain a visualization of so-called Λ-vortices and their breakdown into smaller Ω-shaped vortices during the transition process next to the wall." However, the mechanism of formation of multiple ring-like vortex chains has not been clearly investigated and documented. It is very hard to believe that the "Crow theory" manipulates the multiple ring structure formation according to our DNS observation. First, the ring is perpendicular to the original legs but not parallel to the legs like "Crow theory" and, second, the multiple rings are formed one by one, but not simultaneously like "Crow theory". On the other hand, the mechanism of vortex "breakdown and reconnection" does not follow the Helmoholtz vorticity conservation law. Therefore, the mechanism of multiple ring structure formation is still not very clear and open for research. The purpose of this paper is to give a detailed description on physical process and deeper analysis for the mechanism of multiple ring structure formation. In order to get deep understanding the mechanism of the late flow transition in a boundary layer and physics of turbulence, we recently conducted a high order direct numerical simulation (DNS) with 1920×241×128 gird points and about 600,000 time steps to study the mechanism of the late stages of flow transition in a boundary layer at a free stream Mach number 0.5 (Chen et al., 2009, 2010a, 2010b, 2011a, 2011b; Liu et al., 2010a, 2010b, 2010c, 2011a, 2011b, 2011c, 2011d; Lu et al., 2011a, 2011b, 2011c). The work was supported by AFOSR, UTA, TACC and NSF Teragrid.

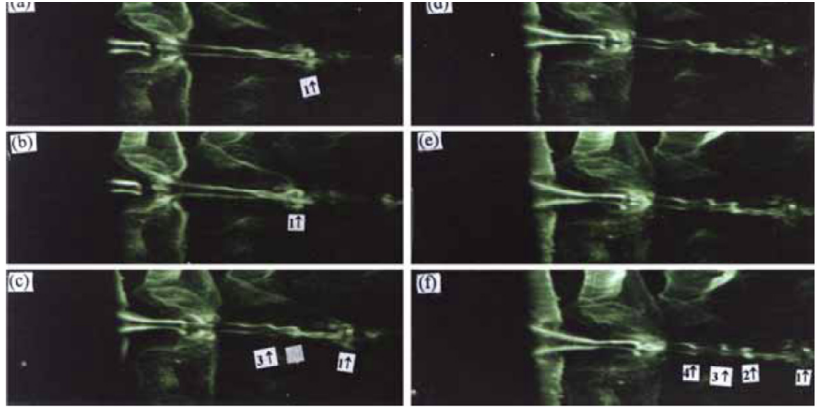

*Figure 1: Evolution of the ring-like vortex chain by experiment (Lee et al, 2007)*

A vortex identification method introduced by Jeong & Hussain (1995) is applied to visualize the vortex structures by using an iso-surface of a $\lambda_2$-eigenvalue. The vortex cores are found by the location of the inflection points of the pressure in a plane perpendicular to the vortex tube axis. The pressure inflection points surround the pressure minimum that occurs in the vicinity of the vortex core. By this $\lambda_2$-eigenvalue visualization method, the multiple ring-like vortex structures shaped by the nonlinear evolution of T-S waves in the transition process are shown in Figure 2 by our DNS result.

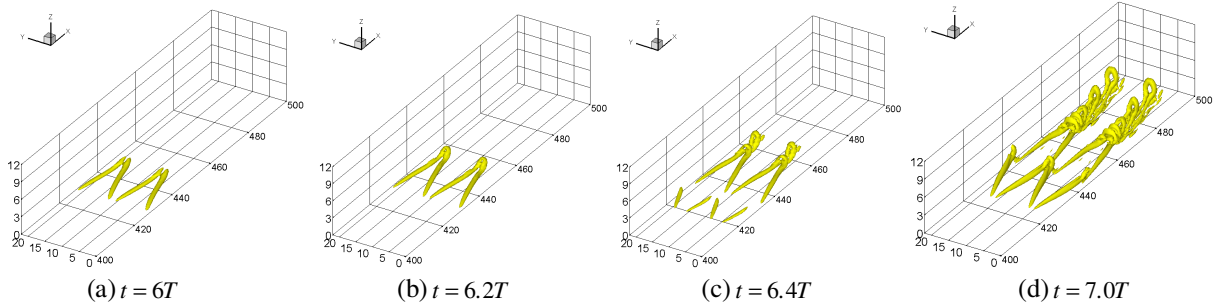

(a) $t = 6T$      (b) $t = 6.2T$      (c) $t = 6.4T$      (d) $t = 7.0T$

*Figure 2: The evolution of vortex structures at the late-stage of transition (Liu et al, 2010)*
*(Where T is the period of T-S wave)*

By a careful observation and analysis, we recently found that there is a momentum deficit area around each ring-like vortex during its formation and stretching. Several questions will be raised. Where does the low speed flow come from? What is the relation between the momentum deficit and the multiple ring-like structures? The existing theory and hypothesis did not provide a clear mechanism about these questions.

## II. Case setup and DNS validation

**2.1 Case setup**

The computational domain is displayed in Figure 3. The grid level is 1920×128×241, representing the number of grids in streamwise (*x*), spanwise (*y*), and wall normal (*z*) directions. The grid is stretched in the normal direction and uniform in the streamwise and spanwise directions. The length of the first grid interval in the normal direction at the entrance is found to be 0.43 in wall units ($y^+ = 0.43$).

The parallel computation is accomplished through the Message Passing Interface (MPI) together with domain decomposition in the streamwise direction. The computational domain is partitioned into N equally-sized sub-domains along the streamwise direction. N is the number of processors used in the parallel computation. The flow parameters, including Mach number, Reynolds number etc are listed in Table 1. Here, $x_{in}$ represents the distance between leading edge and inlet, $Lx$, $Ly$, $Lz_{in}$ are the lengths of the computational domain in x-, y-, and z-directions, respectively, and $T_w$ is the wall temperature.

| $M_\infty$ | $Re$ | $x_{in}$ | $Lx$ | $Ly$ | $Lz_{in}$ | $T_w$ | $T_\infty$ |
|---|---|---|---|---|---|---|---|
| 0.5 | 1000 | $300.79\delta_{in}$ | $798.03\delta_{in}$ | $22\delta_{in}$ | $40\delta_{in}$ | 273.15K | 273.15K |

*Table 1. Flow parameters*

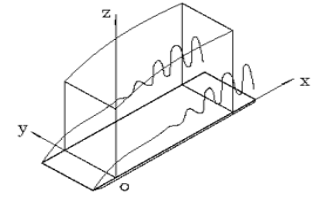

*Figure 3: Computation domain*

### 2.2 Code validation and DNS results visualization
To justify the DNS codes and DNS results, a number of verifications and validations have been conducted.

#### 2.2.1 Comparison with Linear Theory
Figure 4 compares the velocity profile of the T-S wave given by our DNS results to linear theory. Figure 5 is a comparison of the perturbation amplification rate between DNS and LST. The agreement between linear theory and our numerical results is pretty well.

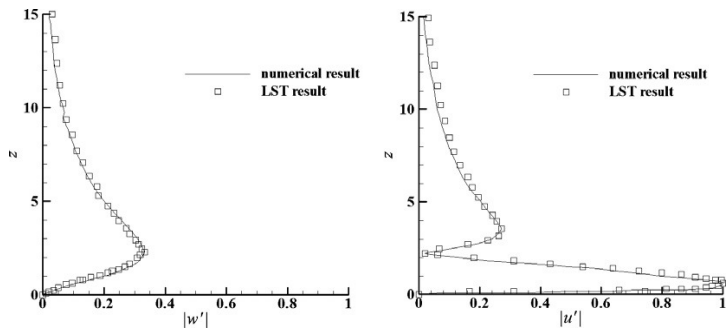
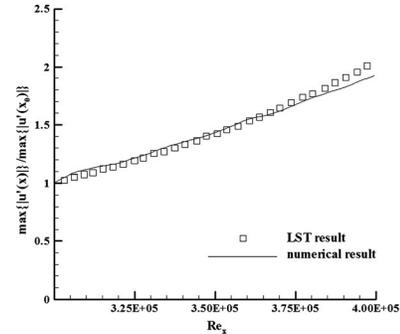

*Figure 4: Comparison of the numerical and LST velocity profiles at Rex=394300*

*Figure 5: Comparison of the perturbation amplification rate between DNS and LST*

#### 2.2.2 Grid Convergence
The skin friction coefficient calculated from the time-averaged and spanwise-averaged profile is displayed in Figure 6. The spatial evolution of skin friction coefficients of laminar flow is also plotted out for comparison. It is observed from these figures that the sharp growth of the skin-friction coefficient occurs after $x \approx 450\delta_{in}$, which is defined as the 'onset point'. The skin friction coefficient after transition is in good agreement with the flat-plate theory of turbulent boundary layer by Cousteix in 1989 (Ducros, 1996). Figure 6 also shows that we get grid convergence in velocity profile.

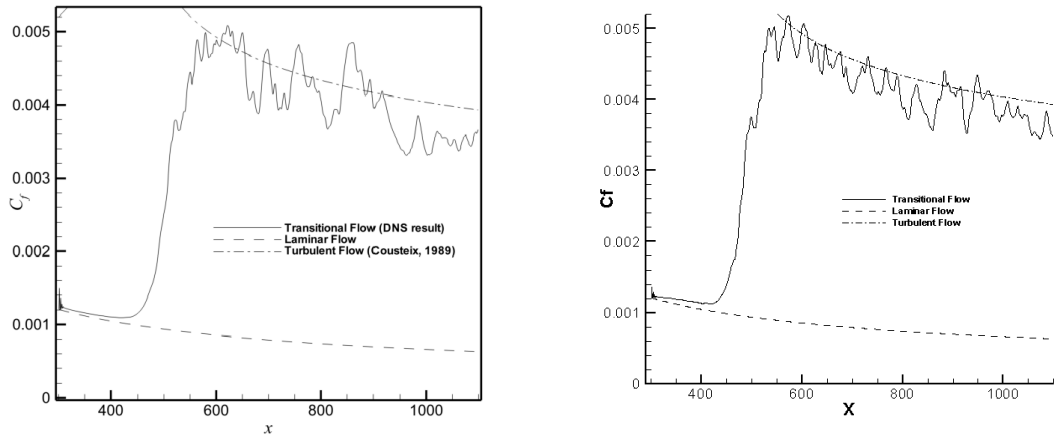

(a) Coarse Grids ($960\times64\times121$)    (b) Fine Grids (1920x128x241)

*Figure 6: Streamwise evolutions of the time-and spanwise-averaged skin-friction coefficient*

### 2.2.3 Comparison with Log Law

Time-averaged and spanwise-averaged streamwise velocity profiles for various streamwise locations in two different grid levels are shown in Figure 7. The inflow velocity profiles at $x = 300.79\delta_{in}$ is a typical laminar flow velocity profile. At $x = 632.33\delta_{in}$, the mean velocity profile approaches a turbulent flow velocity profile (Log law). This comparison shows that the velocity profile from the DNS results is turbulent flow velocity profile and the grid convergence has been realized.

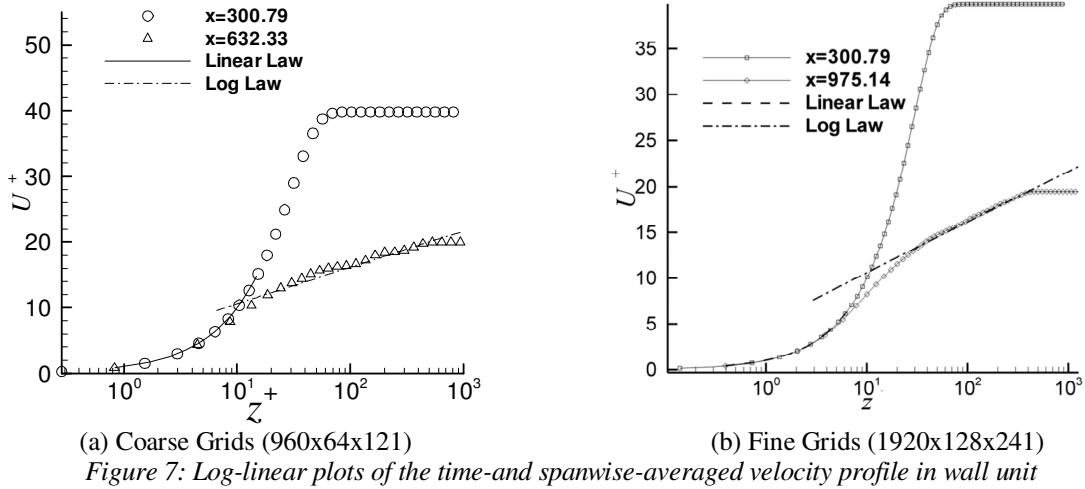

(a) Coarse Grids (960x64x121)     (b) Fine Grids (1920x128x241)
*Figure 7: Log-linear plots of the time-and spanwise-averaged velocity profile in wall unit*

**All these verifications and validations above show that our code is correct and our DNS results are reliable.**

### 2.2.4 Visualization using $\lambda_2$

In boundary layer flows, viscosity is non-negligible. So standard approaches, such as integrating vortex lines, using minimum pressure or maximum vorticity, may lead to improper vortex identification. Jeong and Hussain have established a robust criterion for identification of vortex (or coherent) structures in viscous flows based on the eigenvalues of the symmetric $3\times 3$ tensor

$$M_{ij} := \sum_{k=1}^{3} \Omega_{ik}\Omega_{kj} + S_{ik}S_{kj} ,$$

where

$$\Omega_{ij} := \frac{1}{2}(\frac{\partial u_i}{\partial x_j}+\frac{\partial u_j}{\partial x_i}) \text{ and } S_{ij} := \frac{1}{2}(\frac{\partial u_i}{\partial x_j}-\frac{\partial u_j}{\partial x_i})$$

represent the symmetric and anti-symmetric components of the velocity gradient tensor, $\nabla u$. Given the three (real) eigenvalues of M at each grid point, a vortex core is identified as any contiguous region having two negative eigenvalues. If the eigenvalues are sorted such that $\lambda_1 \leq \lambda_2 \leq \lambda_3$, then any region for which $\lambda_2 < 0$ corresponds to a vortex core. One advantage of this approach is that vortices can be identified as isosurfaces of a well-defined scalar field. Moreover, the criterion $\lambda_2(x) < 0$ is scale invariant, so in principle there is no ambiguity in selecting which isosurface value to render. In practice, one usually biases the isosurface to a value that is below zero by a small fraction of the full dynamic range in order to avoid noise in regions where the velocity is close to zero. Here, we use the $\lambda_2$ criterion (Jeong & Hussain, 1995) for visualization.

### III. Mechanism of Momentum deficit

The flow field around each ring-like vortex and surrounding area has been studied in details. Further more, 3-D structure of the positive spike, negative spike, shear layer are also obtained by our numerical simulation. The ring-like vortices have been found that they are actually formed one by one along the streamwise direction (figure 8) and they are very stable travelling down stream for a long distance. In order to study the 3-D structures of the vortices, 4

4
American Institute of Aeronautics and Astronautics

cross-sections are selected for analysis with streamwise velocity contour as background at Fig.9. From the observation of figure 10, we can note that there is a momentum deficit area generated on each slice which looks like a circle. They are initially located in a lower position, but then gradually lift up when propagating downstream, which is due to the boundary layer mean velocity profile. Another phenomenon is that the strength of the momentum deficit becomes weak when moving downstream.

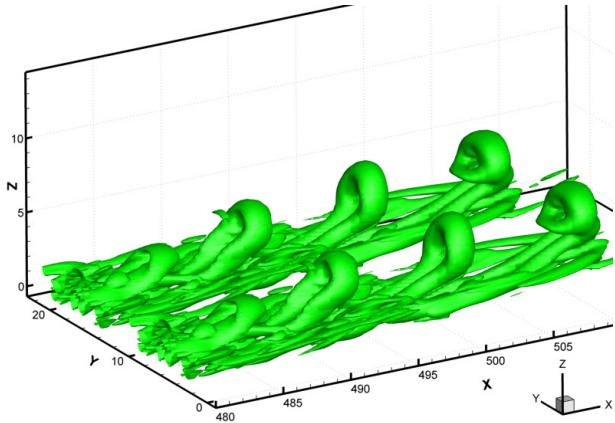 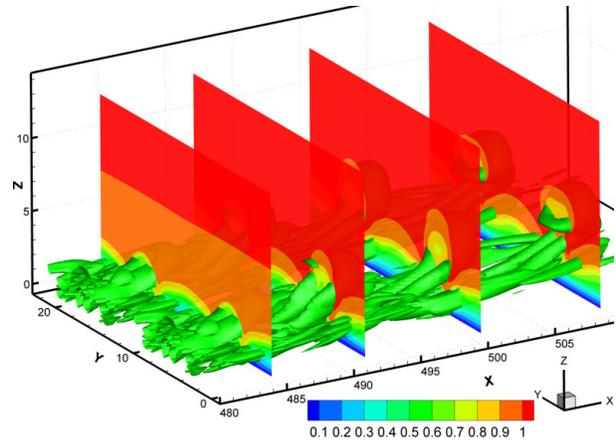

Figure 8: *Isosurface of* $\lambda_2$     Figure 9: *Isosurface of* $\lambda_2$ *and velocity distribution*

A strong momentum deficit has been found around ring-like vortex which is located above the primary vortex legs. It causes a strong shear layer as shown in Fig.10. For clarifying the mechanism, the typical structure of the deficit is shown again with velocity vector (Fig.11) and streamtrace distribution (Fig.12). Inside the deficit area, there are two counter-rotating primary vortices which bring low speed flow from the near wall region (lower boundary layer) to the inviscid area. This phenomenon is caused by vortex ejection, which is induced by the rotation of ring-like vortex and primary vortex legs. From Fig.13, by drawing 3-dimensional vector around ring-like structure, it can be clearly seen that a strong motion of ejection is induced from the center of two ring-like vortex legs to the top area where a momentum deficit is formed.

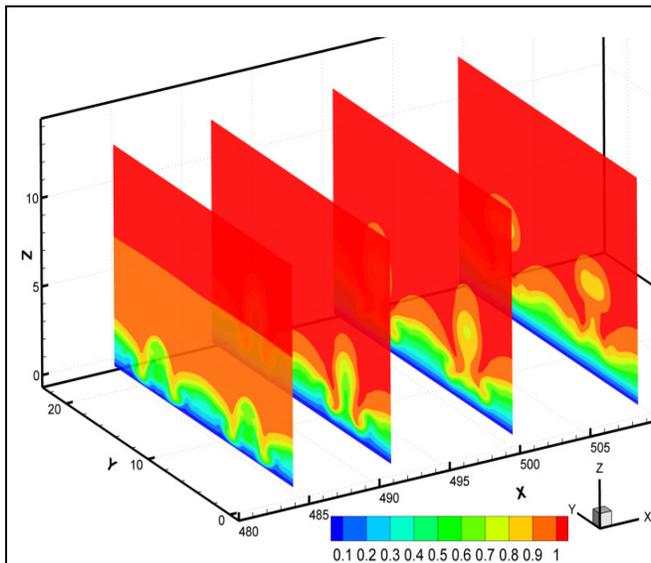 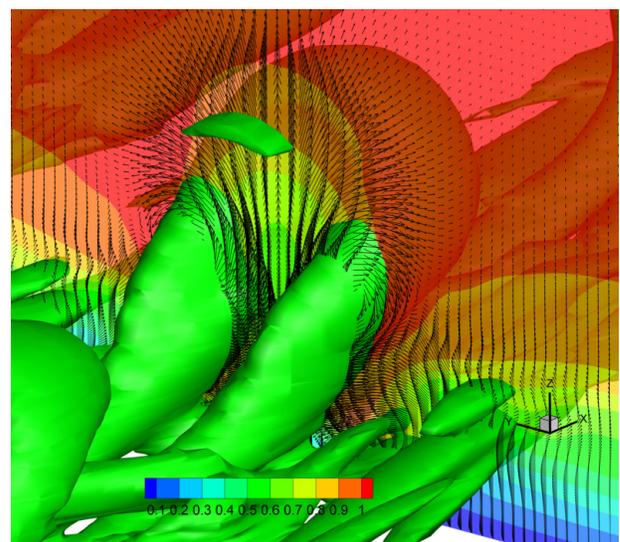

Figure 10: *Four cross-sections with velocity distribution*     Figure 11: *velocity and vector distribution*

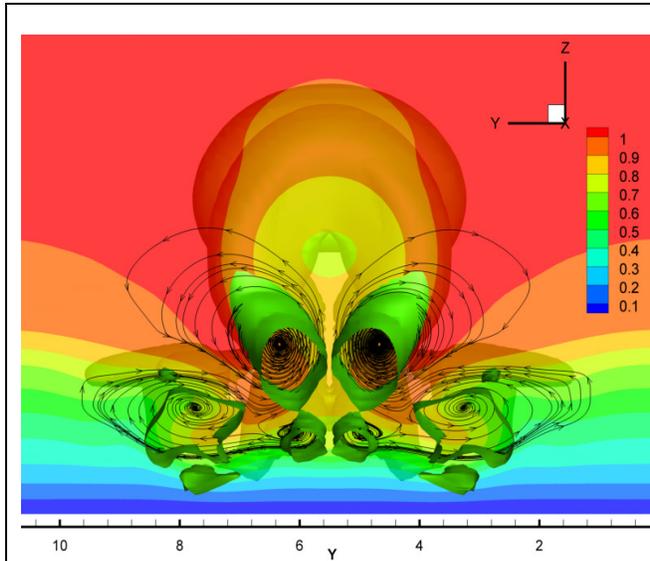

*Firgure 12: velocity and streamtrace distribution*

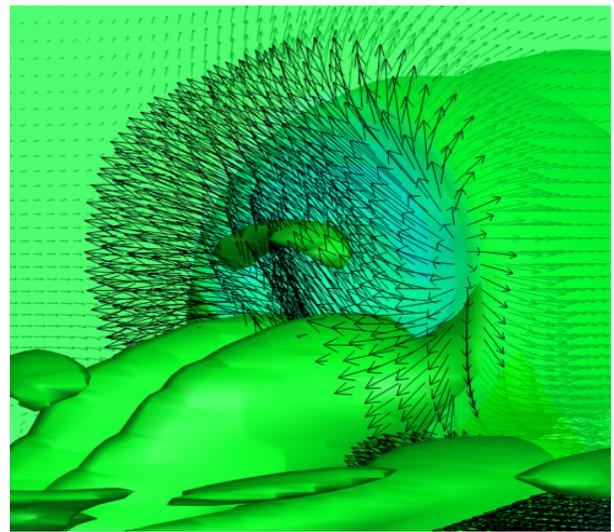

*Figure 13: Iso-surface of $\lambda_2$ and 3-D vector distribution(blue area – momentum deficit)*

To further validate the above conclusion about where the slow fluid mainly comes from, the cross-section of spanwise with vector distribution is drawn at the same time step as shown in Fig. 12. The results show that there are low speed area above the primary vortex legs and outside the ring-like structures where a momentum deficit is formed due to the ejection motion, which means the main source of the momentum deficit is convected from the lower boundary layer. As a consequence, a shear layer region is formed around the momentum deficit. Meanwhile, a corresponding high speed region is generated underneath the ring-like vortex which is caused by the sweep motion which bring high speed flow from inviscid area down to lower boundary layer (Fig 15). Another question will be raised that what kind of mechanism causes the shear layer instability for multiple ring formation. Let us first look at the instability of shear layer and then study how the shear layer becomes vortex at the inviscid area due to shear layer instability.

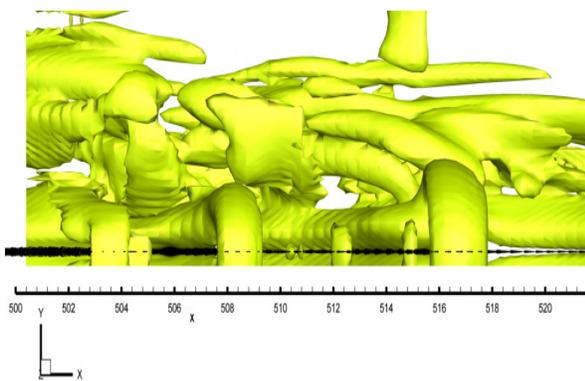

*Figure 14: top view Isosurface of $\lambda_2$*

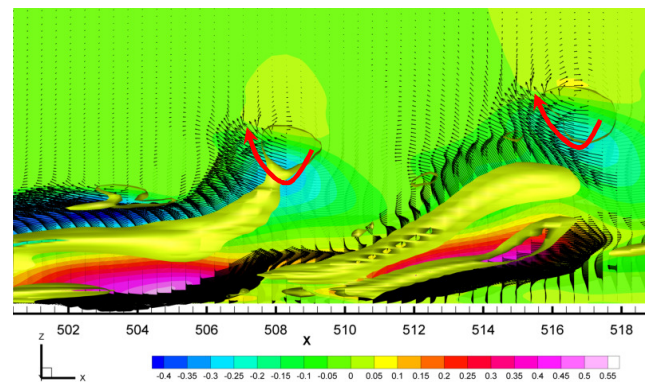

*Figure 15: side view of velocity and vector distribution*

**IV. 2-D Kelvin-Helmoholz instability**

2-D instability caused by shear layer with inflow disturbances and formation of pairing vortex rings were obtained by our previous calculation. It is usually called Kelvin-Helmholtz instability which is inviscid instability.

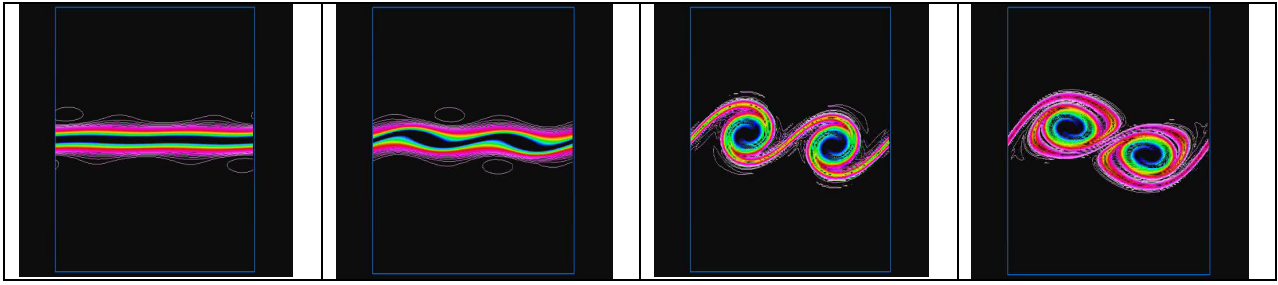

Figure 16: Numerical simulation on Kelvin-Helmoholtz instability

## V. Shear layer instability analysis

The distributions of averaged streamwise-velocity are given in Fig. 17a along the normal grid lines at the center plane. The streamwise positions of the line is x=491.1 $\delta_{in}$. The dip of the line corresponds to the momentum deficit. It can be seen clearly that there a high shear layer in the central plane which is located above the ring legs. The second order derivative $\partial^2 U/\partial z^2$ is calculated to demonstrate the existence of the inflection points. The existence and correspondence of the inflection point at the upper is illustrated by two dashed lines intersecting the distribution of the streamwise velocity and its second order derivative (Fig. 17b).

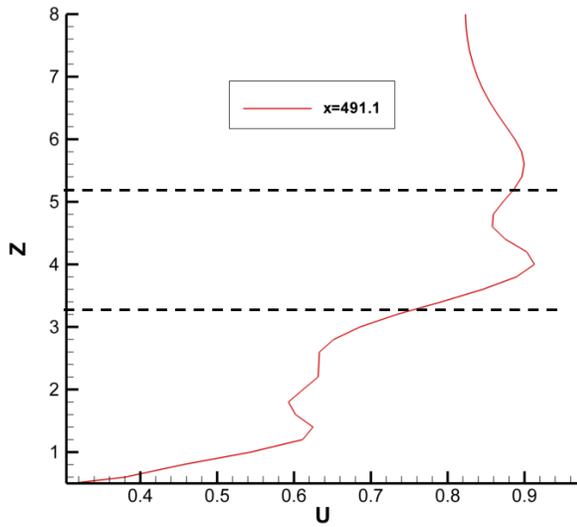
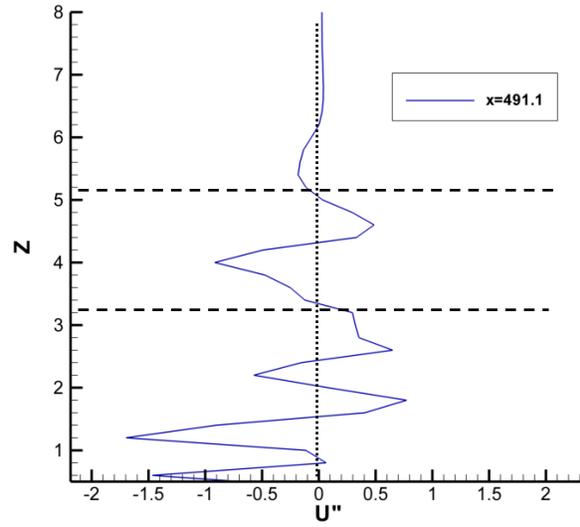

Figure 17a: The streamwise velocity distribution    Figure 17b: The distribution of second order derivative of corresponding streamwise velocity

It is obvious that the existence of the inflection points (e.g. z=5.2) in shear layers will cause the flow instability and generates vortex rollers according to the shear layer instability theory in 2-D. So the mechanism for the vortex ring generation is closely related to the shear layer instability. Loss of the stability of the shear layer will result in the formation of the vortex rings for 3-D flow.

### 5.1 Governing equations and numerical methods for linear shear layer stability

For shear layer stability analysis, we only consider the 2-D incompressible flow here. For non-dimensional incompressible NS eqs:

$$\begin{cases} \dfrac{\partial V}{\partial t} + V \cdot \nabla V = -\nabla p + \dfrac{1}{\text{Re}} \nabla^2 V \\ \nabla \cdot V = 0 \end{cases} \quad (1)$$

Where, $V = (u, v)$

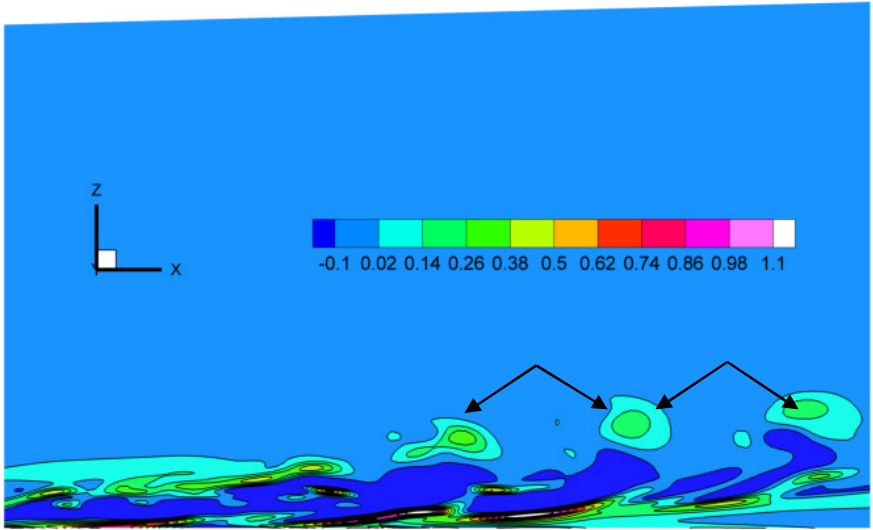

*Figure 18. Scalar field of the gradient of velocity*

For our problem, the spatial disturbance is easy to measure. Actually, it is related to the distance among two neighboring vortices in the central x-z plane (Fig.18).

Introduce disturbed stream function: $\Psi(x, y, t) = \psi(y) e^{i\alpha(x-ct)}$, and the disturbed velocities as $u'$, $v'$, then the original velocities will be

$$\begin{cases} u(x, y, t) = U_0(x, y, t) + u'(x, y, t) = U_0(y) + \dfrac{d\psi}{dy} e^{i\alpha(x-ct)} \\ v(x, y, t) = V_0(x, y, t) + v'(x, y, t) = V_0(y) - \dfrac{d\psi}{dx} e^{i\alpha(x-ct)} \end{cases} \quad (2)$$

In which, $U_0(y)$ is the streamwise averaged velocity and the averaged vertical velocity $V_0(y) = 0$.

If temporal mode is considered to the stability problem, we can make the wave number $\alpha$ as a real number and $c = c_r + i c_i$ as a complex number.

By plugging in eq (2) to (1) and canceling the pressure p, we can get the linearized stability O-S equation:

$$[(U_0 - c)(D^2 - \alpha^2) - D^2 U_0]\psi = -\dfrac{i}{\alpha \text{Re}}(D^2 - \alpha^2)\psi \quad (3)$$

And $D = \dfrac{d}{dy}$

**5.2 Finite difference method (central difference scheme)**

$$\begin{cases} D^2 \psi = (\psi_{n+1} - 2\psi_n + \psi_{n-1})/h^2 \\ D^4 \psi = (\psi_{n+2} - 4\psi_{n+1} + 6\psi_n - 4\psi_{n-1} + \psi_{n-2})/h^2 \end{cases} \quad (4)$$

## 5.3 Numerical method of O-S eq

Coefficient $c_i$ determines the property of stability and the flow will be unstable if $c_i > 0$ and stable if $c_i < 0$. Apply (4) to (3) we can get the generalized eigenvalue problem:

$$A\varphi + Bc\varphi = 0 \tag{5}$$

Where A and B are the coefficients' matrix and the general eigenvector $\varphi$ denotes the values of $\psi$ at different position. $c = c_r + ic_i$ will be the generalized eigenvalue of eq (5).

The second order central difference scheme is used to get the finite different equation from equation (3), then a so called eigenvalue method is applied to get the value for $c$ which should be a complex number. We can get the physical solution of the frequency $c$, whose imaginary part $c_i$ is about 0.045 for our case. The positive value of $c_i$ means the shear layer is unstable. Fig.19 shows the corresponding shape of the corresponding eigenvector function $\psi(y)$.

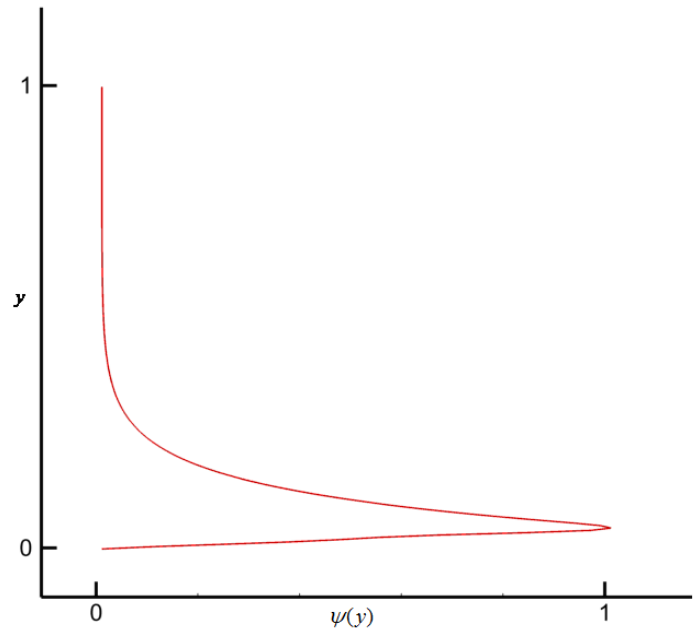

*Figure 19. Shape function*

### VI. Mechanism of multiple rings formation

In order to explore the mechanism of the formation of multiple vortex rings, the 3-Dimensional distributions of streamwise-velocity are given in Fig. 20(a) along the streamwise direction where green color represents the negative spike (momentum deficit) around each ring-like vortex.

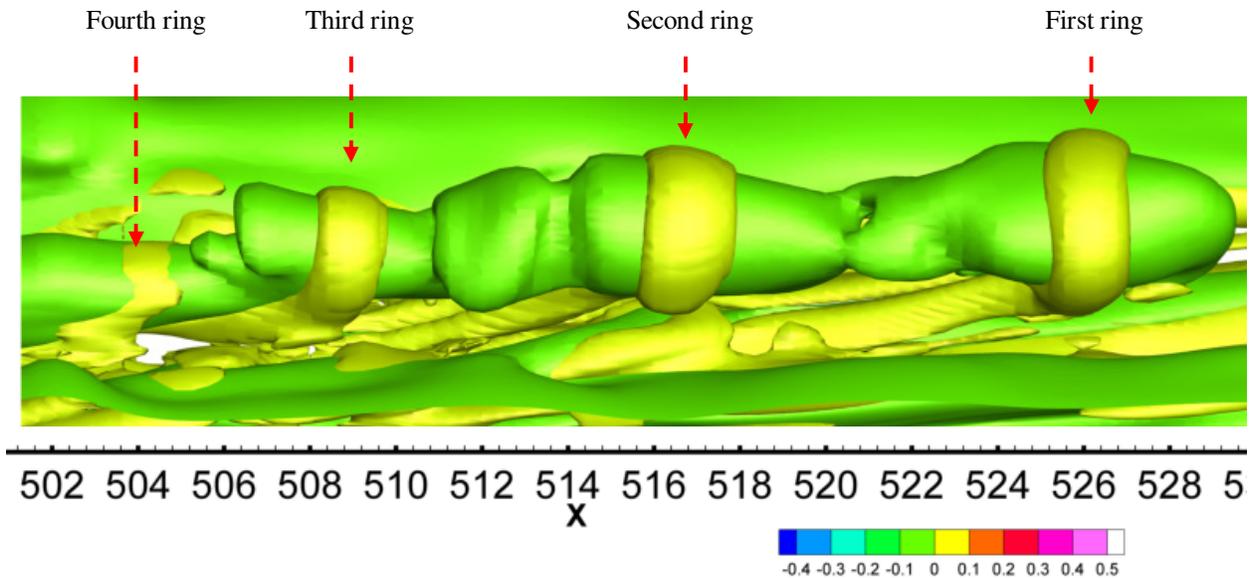

*Figure 20(a). 3-D momentum deficit and Ring-like vortices*

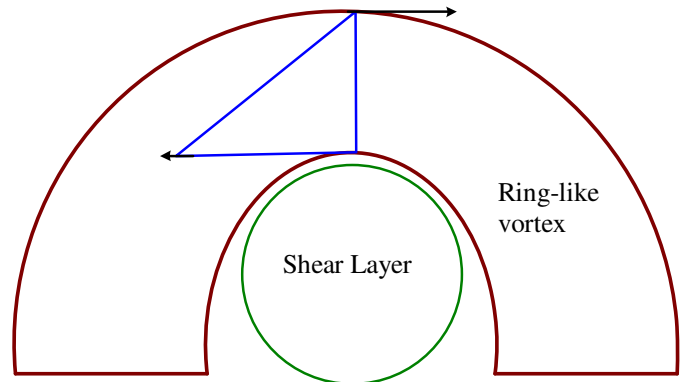

*Fig 20(b). Sketch of mechanism of ring-like vortex formation*

According to our computation, we find that the formation of multiple ring-like vortices is caused by the instability of shear layer discussed above. Originally, the bottom part of the ring is located at viscous area where the flow is stable, however, the top of the ring is located at the inviscid area ( Fig 21(a). at x=508). Due to the instablity of high shear layer near the inviscid area, a ring-like vortex is formed with two legs connected with primary vortex, while the head of the ring-like vortex is lifted up to a higher position inside the inviscid area, a whole ring-like vortex (from the Fig 21(b). at x= 536) is formed with a neck connected to the primary legs. Meanwhile, we illustrate that the multiple rings are formed one by one but not simultaneously. Due to the mean velocity profile, the top of flow which is gradually lifted up to the inviscid area while going downstream. By Fig. 20(a), we can find that the height of the four different rings is increasing.

Since the total vorticity keeps the same inside the flow according to the Helmoholtz vorticity conservation, the streamwise and spanwise vorticity is alternating from each other during the multiple ring formation process. In order to investigate the relationship between the vortices, we measured the distribution of the maximum value of both streamwise and spanwise vorticities in a specific region to avoid the affection of the viscous sub-layer where the vorticity is very large due to the wall surface. Fig.22 is corresponding to Fig.21 for the same x-position. By looking at streamwise and spanwise vorticity distribution at x=508, we can see that at the place, where ring-like vortex is formed, the spanwise vorticity is increased but the streamwise viorticity is sharply decreased. That is because the vorticity is transferred from the streamwise direction to the spanwise direction. We also find that the rotation of the streamwise vortices becomes weaker, however, the rotation of spanwise vortices becomes stronger when the ring-like vortex forms, which can be seen in Fig.21.

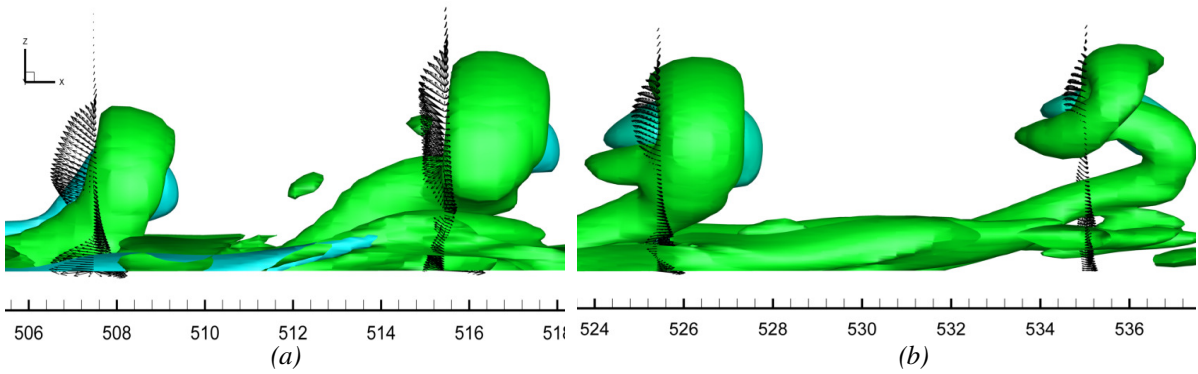

*Figure 21. Ring-like vortex and 3-D vector distribution*

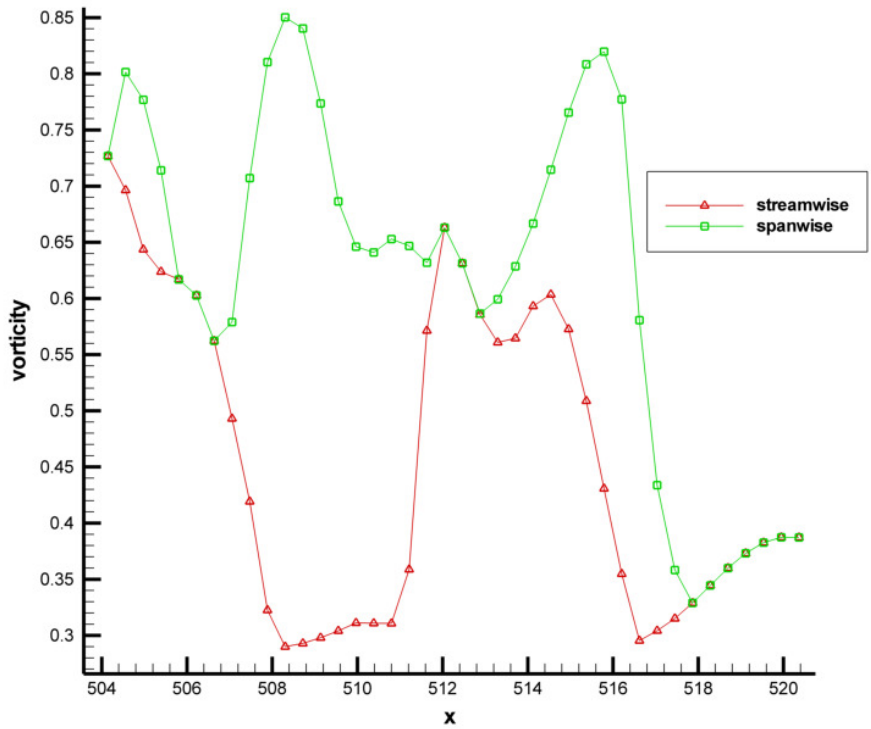

*Figure 22. Vorticity distribution*

We have run another case for microramp vortex generator (MVG) at Mach 2.5 which is shown at Fig 23. We selected three cross-sections (Fig. 24) along streamline by investigating the momentum deficit and concluded that the instability of shear layer is the mechanism that vortex rings are generated. Meanwhile, it proves that these two cases have the same mechanism of multiple rings formation no matter for high speed or low speed flows.

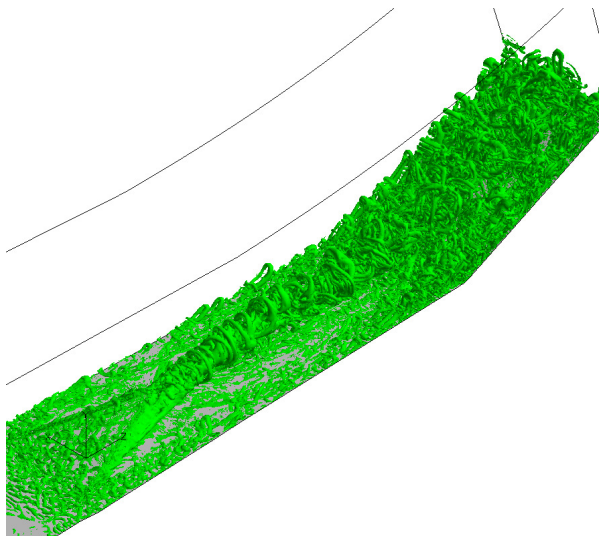
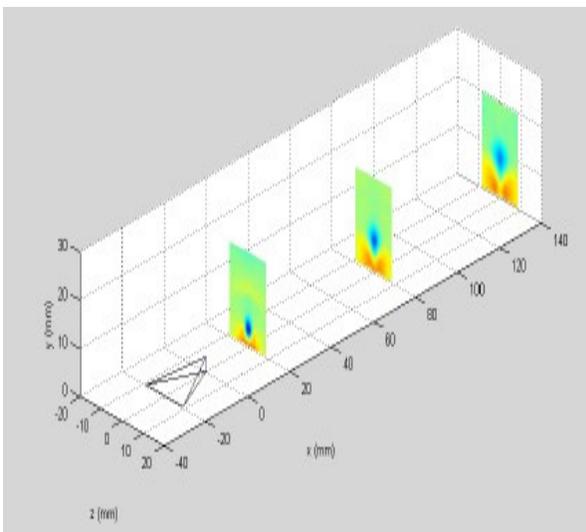

*Fig. 23 Isosurface of $\lambda_2$*   *Fig. 24 Momentum deficit shown in different streamwise directions*

In order to clearly illustrate the mechanism of multiple rings formation and small length scales generation which we have published paper before, we made sketch for a 3-D structure in Fig 25.

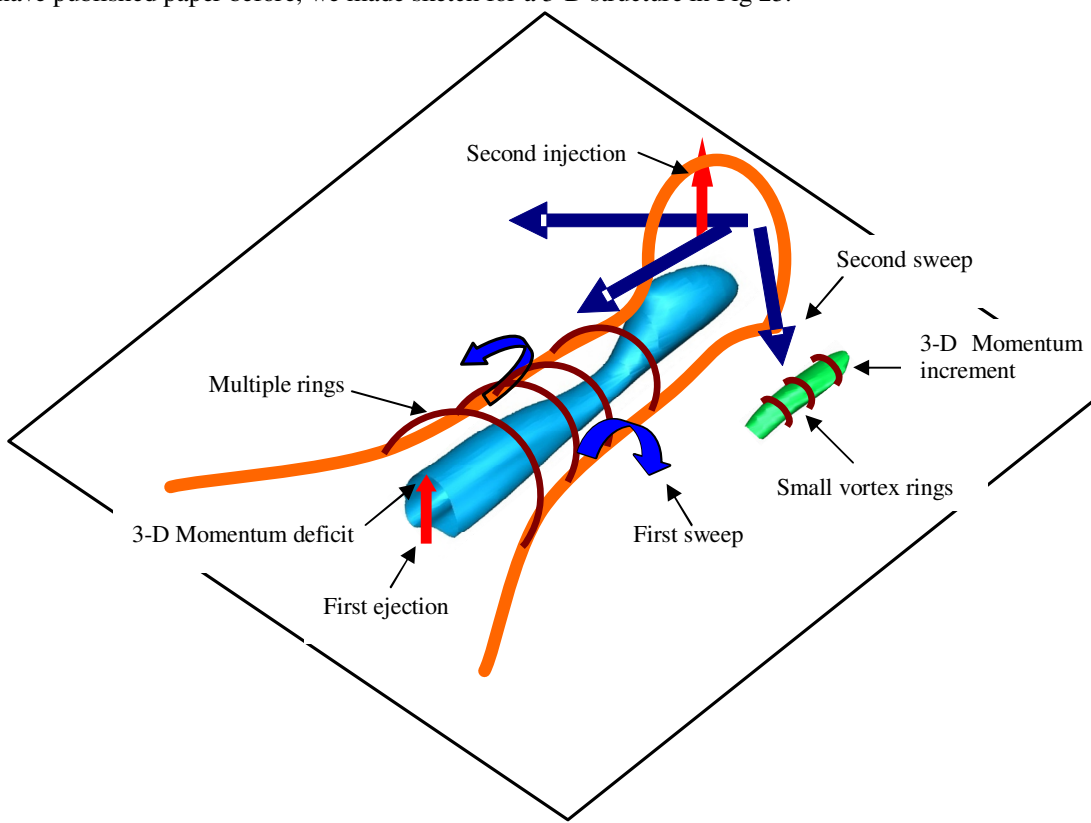

Fig 25. Sketch of mechanism of multiple rings formation and small vortices formation

## VII. Conclusion

The multiple rings formation has been observed by both experiment and DNS. By analysis of the data obtained by our DNS, the following conclusion can be made:
1. The multiple rings are formed one by one but not simultaneously.

2. The momentum deficit is caused by vortex ejection around the original vortex ring legs which brings low speed flow from lower boundary layer to upper region.
3. The shear layer formed by momentum deficit is not stable near the inviscid area due to the inflection points.
4. The ring-like vortex is formed when it is lifted up to the inviscid area where the shear layer is unstable. So, we conclude that the instability of shear layer is the mechanism of multiple rings formation.
5. Investigation of vorticity distribution along streamwise direction reveals the fact that part of streamwise vorticity transports to spanwise vorticity to form multiple ring-like vortices.

**References**

[1] Bake S, Meyer D, Rist U. Turbulence mechanism in Klebanoff transition:a quantitative comparison of experiment and direct numerial simulation. J.Fluid Mech, 2002 , 459:217-243
[2] Borodulin V I, Gaponenko V R, Kachanov Y S, et al. Late-stage transition boundary-Layer structure: direct numerical simulation and exerperiment. Theoret.Comput.Fluid Dynamics, 2002,15:317-337.
[3 ] Chen, L., Liu, X., Tang, D., and Oliveira M., Liu, C., Evolution in the ring-like vortices and spike structure in transitional boundary, Science of China, Physics, Mechanics & Astronomy, Vol.53 No.3: pp514-520, March, 2010a
[4] Chen, L., Liu, X., Oliveira, M., Liu, C., DNS for Ring - Like Vortices Formation and Roles in Positive Spikes Formation AIAA Paper Number 2010-1471, 2010b
[5]Chen, L and Liu, C., Numerical Study on Mechanisms of Second Sweep and Positive Spikes in Transitional Flow on a Flat Plate, Journal of Computers and Fluids, Vol 40, p28-41, 2010c
[6] Crow S C. Stability theory for a pair of trailing vortices. AIAA J, 1970, 8: 2172-2127
[7]Duros F, Comte P, Lesieur M. Large-eddy simulation of transition to turbulence in a boundary layer developing spatially over a flat plate. J.Fluid Mech, 1996, 326:1-36
[8]Guo, Ha; Borodulin, V.I.; Kachanov, Y.s.; Pan, C; Wang, J.J.; Lian, X.Q.; Wang, S.F., Nature of sweep and ejection events in transitional and turbulent boundary layers, J of Turbulence, August, 2010
[9] Jeong J., Hussain F. On the identification of a vortex, J. Fluid Mech. 1995, 285:69-94
[10] Kachanov, Y.S. On a universal mechanism of turbulence production in wall shear flows. In: Notes on Numerical Fluid Mechanics and Multidisciplinary Design. Vol. 86. Recent Results in Laminar-Turbulent Transition. — Berlin: Springer, 2003, pp. 1–12.
[11] Kloker, M and Rist U. , Direct Numerical Simulation of Laminar-Turbulent Transition in a Flat-Plate Boundary Layer on line at http://www.iag.uni-stuttgart.de/people/ulrich.rist/publications.html, 2011
[12] Lee C B., Li R Q. A dominant structure in turbulent production of boundary layer transition. Journal of Turbulence, 2007, Volume 8, N 55
[13] Liu, C. and Chen, L., DNS for Late Stage Structure of Flow Transition on a Flat-Plate Boundary Layer AIAA Paper Number 2010-1470, 2010a, to appear in Journal of Computers and Fluids.
[14] Liu, X., Chen, Z. and Liu, C., Late-Stage Vortical Structures and Eddy Motions in a Transitional Boundary Layer, CHIN. PHYS. LETT. Vol. 27, No. 2 (2010) 024706, 2010b
[15] Liu, C. and Chen, L., Parallel DNS for Vortex Structure of Late Stages of Flow Transition, Journal of Computers and Fluids, available online at: http://dx.doi.org/10.1016/j.compfluid.2010.11.006, 2010d
[16] Lu, P., Liu, C., Numerical Study of Mechanism of Small Vortex Generation in Boundary Layer Transition, AIAA Paper, 2011-0287, January 2011a
[17] Lu, P., Liu, C., Numerical Study of Mechanism of U-Shaped Vortex Formation, AIAA Paper 2011-0286, January 2011b
[18] Lu, P., Liu, C., DNS study on mechanism of small length scale generation in late boundary layer transition. Physica D: Nonlinear Phenomena, Volume 241, Issue 1, 1 January 2012c
[18] Rist, U., Kachanov, Y.S.: Numerical and experimental investigation of the K-regime of boundary-layer transition. In: Kobayashi, R. (ed.) Laminar-Turbulent Transition, pp. 405-412. Springer, Berlin (1995)